\documentclass[%
superscriptaddress,
preprint,
 amsmath,amssymb,
 aps,
pra,
showpacs,
floatfix,
]{revtex4-1}

\usepackage{graphicx}
\usepackage{dcolumn}
\usepackage{bm}
\usepackage[english]{babel}

\begin{document}
\thispagestyle{empty}

\title{High Harmonic Generation with Orbital Angular Momentum Beams: Beyond-dipole Corrections}

\author{Esra Ilke Albar}
 \email{esra-ilke.albar@mpsd.mpg.de}
 \affiliation{Max Planck Institute for the Structure and Dynamics of Matter,\\ Center for Free Electron Laser Science,\\ Luruper Chaussee 149, 22761 Hamburg, Germany}
\author{Valeriia P. Kosheleva}%
 \affiliation{Max Planck Institute for the Structure and Dynamics of Matter,\\ Center for Free Electron Laser Science,\\ Luruper Chaussee 149, 22761 Hamburg, Germany}
\author{Heiko Appel}
 \affiliation{Max Planck Institute for the Structure and Dynamics of Matter,\\ Center for Free Electron Laser Science,\\ Luruper Chaussee 149, 22761 Hamburg, Germany}
\author{Angel Rubio}
\affiliation{Max Planck Institute for the Structure and Dynamics of Matter,\\ Center for Free Electron Laser Science,\\ Luruper Chaussee 149, 22761 Hamburg, Germany}
\affiliation{Center for Computational Quantum Physics (CCQ), The Flatiron Institute, 162 Fifth Avenue, New York, New York 10010, USA}
\author{Franco P. Bonafé} 
\email{franco.bonafe@mpsd.mpg.de}
\affiliation{Max Planck Institute for the Structure and Dynamics of Matter,\\ Center for Free Electron Laser Science,\\ Luruper Chaussee 149, 22761 Hamburg, Germany}


\begin{abstract}
We study the high harmonic generation with vortex beams beyond the dipole approximation.
To do so we employ the full minimal coupling approach to account for multipolar coupling without truncation and describe the full spatio-temporal properties of the electromagnetic field.
This allows us to investigate the beyond-dipole deviations in electron trajectories and the emitted power, where the influence of the orbital angular momentum contains both magnetic and quadrupolar effects. 
In contrast to the system driven by plane-wave light, we show that the non-linear dipole dynamics induced by the vortex beams are not confined to the polarization or propagation directions, but also have a component in the orthogonal direction. We identify the effects of the resulting symmetry breaking via increased beyond dipole corrections which are particularly apparent in even harmonics.
\end{abstract}

\maketitle

\section{\label{sec:intro} Introduction}
The fact that light can carry orbital angular momentum (OAM) was first theoretically predicted by Allen et al. \cite{Allen1992} and experimentally realized by He et al. \cite{He1995}. Nowadays, OAM beams (vortex or twisted-light beams) are routinely produced using spiral phase plates \cite{Sueda:04}, forked holograms \cite{Heckenberg:92}, q-plates \cite{Karimi2009}, metallic nanostructures \cite{Prinz2023, Kerber2018, Albar2023}, and many others (for more details, see reviews \cite{Chen2020,Xinyuan2021}).  
Due to OAM twisted light has found applications in optical manipulation \cite{Grier2003,Min2013}, quantum information \cite{Magana-Loaiza_2019,PhysRevApplied.11.064058}, imaging and microscopy \cite{Tamburini2006,Maurer2010}, super-resolution optical sensing \cite{Drechsler2021}, and many others. 
Numerous significant theoretical and experimental investigations have been conducted to gain a deeper understanding of the interaction between twisted light and matter, encompassing processes such as excitation \cite{SolyanikGorgone2019, Schmiegelow2016,Fuks2023,PhysRevA.88.033841,Afanasev2018,PhysRevLett.119.253203,Begin2023,PhysRevA.107.023106,Lange2022,Peshkov2023}, ionization \cite{Kosheleva2020,Picon2010,PhysRevA.101.033412,PhysRevA.107.033112,Kiselev2023,Kiselev2024,Kaneyasu2017,Ninno2020}, scattering \cite{Peshkov2018,Peshkov2019,Serbo2022,Forbes2019}, and various others.\\
\noindent
Recently, the study of high harmonic generation (HHG) with twisted light has gained momentum \cite{photonics4020028,Toda:10,zurch2012,PhysRevLett.113.153901,Gauthier2017,doi:10.1021/acsphotonics.1c01768,Hernandez-Garcia2013,Geneaux2016,Kong2017,Sanson:18,Rego2022,Rego2019}, as it offers a tabletop setup for generating OAM-carrying light beams. 
The HHG is understood as a three-step model \cite{Lewenstein1994}: I) the electron is ionized from a parent atom, then II) it propagates in the laser field and finally III) it recombines with a parent atom and emits high harmonics. For optical lasers with intensities around $5.0$ V/nm, a typical electron motion range is much smaller than the wavelength of a driving laser. Therefore, the spatial profile of the laser field in step II) is neglected, i.e. the electric dipole approximation is applied and the electron is assumed to be driven solely by an electric field.
However, for high-speed electrons their trajectories can be significantly influenced by the magnetic field component of the Lorentz force, resulting in a figure-eight electron motion \cite{Maurer2021, Walser2000, Forre2006, Madsen2022, Suster2023, Vendelbo2020, Jensen2022}.

Moreover, twisted-light beams have a spatial-dependent intensity profile, which causes the usual dipole approximation to fail to describe the interaction between matter and these types of beams. So far, HHG with OAM beams has been studied within local dipole approximation \cite{Hernandez-Garcia2013,Rego2016}, which merely accounts for local values of the field. In this work, in contrast, we consider the coupling between light and matter beyond dipole approximation, allowing us to account for the effects of field gradients and magnetic fields. To investigate these effects, we perform real-space and real-time numerical simulations and inspect the HHG spectrum resulting from different matter positions with respect to the beam center. We observe symmetry breaking due to the magnetic field and field gradients which results in the appearance of even harmonics. Moreover, we resolve the electron trajectories and identify the effect of orbital angular momentum. Our analysis shows that the motion of the driven electron is not confined to the polarization or propagation axes only. 
Furthermore, we investigate the angular dependence of the emission of harmonics in the plane orthogonal to the laser propagation, where the off-axis (perpendicular to laser polarization) component shows the largest corrections beyond the dipole level. 
Finally, we show that the yield of such off-axis harmonics can be enhanced by tuning the OAM number of the incident beam.
\\
\noindent
The paper is organized as follows: In Sec. \ref{sec:method} we present the general expression for the high harmonics yield while providing an overview of our beyond dipole approach.
In  Sec. \ref{sec:results} we investigate the beyond dipole effects in high harmonics yield for a case of Bessel beam and compare it to the case of HHG driven by conventional plane wave pulse.
We also address the influence of orbital angular momentum on the HHG process.
Finally, a summary and an outlook
are given in Sec. IV.
The SI system of units is used throughout the paper unless specified otherwise.

\section{\label{sec:method} Method}

We consider the HHG from the hydrogen atom driven by twisted light beams.  We employ a cylindrical simulation box with radius of $1.32$ nm and length of $3.7$ nm. The spacing is set to $0.016$ nm while the timestep is $0.0036$ fs. We propagate the electronic system in real-time utilizing the time-dependent Schrödinger equation 
\begin{equation}
    i\hbar \partial_t \varphi(\mathbf{r},t) = \mathcal{H} \varphi(\mathbf{r},t),
\end{equation}
using the real-space real-time Octopus-code \cite{Tancogne-Dejean2020}.
Here $\hbar$ is the reduced Planck constant and $\mathcal{H}$ is the electronic Hamiltonian:
\begin{equation}
       \mathcal{H}= - \frac{\hbar^2\boldsymbol{\nabla}^2}{2m_e}+V_{\text{nuc}}(\mathbf{r}, t) + \mathcal{H}_{\rm int}.
   \label{eq:total_h}
\end{equation}

Here $m_e$ is the mass of the electron, $V_{\text{nuc}}(\mathbf{r}, t)$ is the Coulomb potential of the nucleus, and $\mathcal{H}_{\rm int}$ is interaction Hamiltonian. In this work, we will go beyond the commonly used multipolar expansion and consider the full minimal coupling (FMC) Hamiltonian without truncations \cite{bonafe2024}:
\begin{equation}
       \mathcal{H}_{\rm int}^{\rm f.m.c.} = -\frac{i\hbar |e| }{cm_e} \mathbf{A}(\mathbf{r},t) \cdot \boldsymbol{\nabla} + \frac{e^2}{2m_ec^2}\mathbf{A}^2(\mathbf{r},t).
   \label{eq:fmc_h_int}
\end{equation}

Here $e$ is the electron charge, and $c$ is the speed of light in vacuum.
To discern the beyond-dipole effects, we also compare the results obtained with light-matter coupling in the electric dipole approximation in the velocity gauge:
\begin{equation}
       \mathcal{H}_{\rm int}^{\rm d.} = -\frac{i\hbar |e| }{cm_e} \mathbf{A}(\mathbf{r}_0,t) \cdot \boldsymbol{\nabla} + \frac{e^2}{2m_ec^2}\mathbf{A}^2(\mathbf{r}_0,t),
   \label{eq:d_h_int}
\end{equation}
where $\mathbf{r}_0$ is the coordinate of the center of charge of the system.
In the present work, we restrict ourselves to the case of a Bessel beam to describe the vortex pulse. The vector potential of a linearly-polarized Bessel pulse propagating along the $z$ axis is given by \cite{Schmidt2024} (see Supporting Information for details on its construction).
\begin{eqnarray}
\mathbf{A}^{(\mathrm{tw})}(\mathbf{r},t) \approx A_0 \sin^2 \left( \frac{\pi (z - R_0)}{w} \right) \theta(z - R_1)  \theta(R_0 - z)  J_{m}\left(\varkappa r_{\perp}\right) \cos(k_z z + m\phi - \omega t)\boldsymbol{e}_x,
\label{eq:bessel_beam}
\end{eqnarray}
where $A_0$ and $\omega$ are the amplitude and energy of the beam, respectively, $m$ is the topological charge of the beam, and $\textbf{e}_x = (1,0,0)$ is the polarization. Here $J_{m}\left(\varkappa r_{\perp}\right)$ stands for the Bessel function, and $r_{\perp}, \phi$, and $z$ are cylindrical coordinates, $k_z$ and $\varkappa = \sqrt{(\frac{\omega}{c})^2-k_z^2}$ are longitudinal and transversal components of momentum $\mathbf{k}$, respectively, and $\theta(x)$ is a Heaviside step function. We employ a sin$^2$ envelope expressed in real space, ranging from $R_0$ to $R_1$ along the $z$ axis. To understand the effects of OAM, we compare the results with the HHG calculation employing a plane wave pulse with explicit space-dependence, namely the following:
\begin{eqnarray}
 \mathbf{A}^{(\mathrm{pw})}(\mathbf{r},t) = 
 A_0 \sin^2 \left( \frac{\pi (z - R_0)}{w} \right)  \theta(z - R_1)  \theta(R_0 - z)  \cos(k_z z  - \omega t)\boldsymbol{e}_x. 
\label{plane_wave}   
\end{eqnarray}

The plane wave is arranged to have the same field amplitude, wavelength (800 nm), phase, and envelope as the Bessel beam with $m=1$ at the $\boldsymbol{r} = (r_{\perp}, 45^{\circ},0)$ with $ r_{\perp}=2545$ nm. 
The resulting maximum electric field amplitude is $2.72\times10^{10}$ V/m. The envelope used corresponds to 8 cycles of the pulse (determined by the variable $w= 8\lambda$), yielding a duration of 21.3 fs.

The HHG yield is calculated from the dipolar emission $P^d$ \cite{Jackson2003}, 
\begin{equation}
     P^{d} = \sum_{i=x,y,z}  P_i^d =
    \frac{\mu_0 \omega^4}{12 \pi c}\sum_{i=x,y,z}  \tilde{d_i}^2 (\omega),
\label{eq:dipole_acceleration}
\end{equation}
as well as quadrupolar contribution to the emission $P^Q$  \cite{Gorlach2020, Jackson2003}
\begin{equation}
     P^{Q} = \sum_{i,j = x,y,z}P_{i j}^Q =
    \frac{\mu_0 \omega^6}{1440 \pi c^3}\sum_{i,j = x,y,z}\tilde{Q}_{i j}^2(\omega).
\label{eq:quad_contribution}
\end{equation}
\noindent
Here, $\mu_0$ is the magnetic vacuum permeability, $\tilde{d_i} (\omega)$ is the Fourier transformation of $i$th component of the time-dependent dipole response ${d_i}(t)$,
\begin{equation}
     {d_i}(t) = |e|\langle \varphi(\mathbf{r},t) |r_i|\varphi(\mathbf{r},t) \rangle,
\label{eq:dip_moment}
\end{equation}
\noindent and $\tilde{Q}_{i j}(\omega)$ is the Fourier transform of the time-dependent quadrupole moments $Q_{i j}(t)$ related to directions $i$ and $j$,
\begin{equation}
     Q_{i j}(t) = |e|\langle \varphi(\mathbf{r},t) |\left(3 r_i r_j-r^2 \delta_{i j}\right)|\varphi(\mathbf{r},t) \rangle,
\label{eq:quad_moment}
\end{equation}
where $\delta_{i j}$ is Kronecker delta.

\section{\label{sec:results} Results}

\begin{figure*}
    \includegraphics[scale=0.4 ]{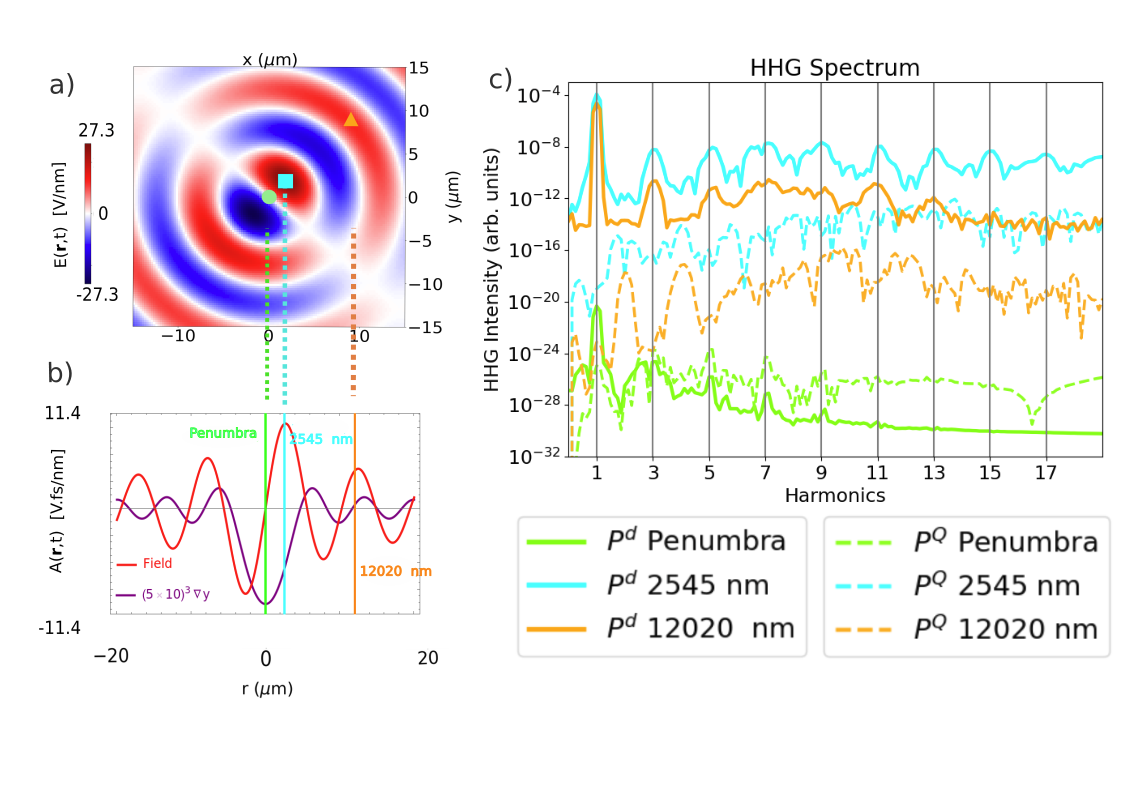}
    \caption{a) Description of the three locations chosen for impact parameter study. The Bessel beam with orbital angular momentum number $m = 1$ has a singularity at the middle of its cross section due to the cancellation of the fields, referred to as penumbra and marked in green. A hydrogen atom is first placed in this position, which results in the absence of driving electric and magnetic fields. However, the maximum of the field gradient occurs here, as seen in panel b). The other locations are the hot spot located in the first ring of the Bessel beam ($x_0 = y_0 =$ 1800 nm, noted by cyan color, with the highest field amplitude and non-zero gradient) and a point in the third ring (12020 nm away with $x_0 = y_0 = 8500$ nm, marked in orange, field amplitude is still considerable, and the gradient is zero). c) The HHG spectra in full minimal coupling from all three locations. Solid lines represent the dipolar emission spectrum $P^d$ (See Eq. \eqref{eq:dipole_acceleration}), while dashed lines refer to the quadrupolar $P^Q$ (See Eq. \eqref{eq:quad_contribution} ) one.
    The presented results are calculated beyond dipole approximation.
    Here and everywhere throughout the paper, the yields below $\approx 10^{-35}$ are considered numerically zero.}
    \label{fig:cross_section}
\end{figure*}

As the electric field of the OAM-beams has a spatially inhomogeneous structure, the HHG yield depends on the atom's position relative to the center of the beam. Therefore, we first investigate this dependence for the case of the Bessel beam with $m=1$. We consider 3 positions of the atom $\boldsymbol{r}_0 = (r_{0}, 45^{\circ},0)$: beam center (penumbra, $r_0 = 0$ nm), first ring (hot spot, $r_0 = 2545$ nm, with $x_0 = y_0 = 1800$ nm), and the third ring ($r_0 = 12020$ nm, with $x_0 = y_0 = 8500$ nm ), as depicted in Fig. \ref{fig:cross_section} panel a). Note, that we also adjust the phase to be the same for all three locations. As one can see from panel b) of Fig. \ref{fig:cross_section} the electric field in the penumbra is zero, while the field gradient is the largest. This results in a quadrupolar $P^{Q}$ emission being comparable to the dipolar $P^{d}$ counterpart, see panel c) of Fig. \ref{fig:cross_section}. We note that since the spectra are calculated within the FMC approach, the dipole emission for the penumbra is nonzero and is caused by the higher order terms in the interaction Hamiltonian $\mathcal{H}_{\rm int}^{\rm f.m.c.}$ \eqref{eq:fmc_h_int}. However, due to the absence of field amplitude, the HHG yield at the penumbra point is very weak and with a low cut-off.

On the other hand, the HHG yield for the atom located at the hot spot is stronger and has a cut-off around the 15th harmonic. While the dipolar emission follows the well-known selection rules and peaks only at odd harmonics, quadrupolar emission peaks at the even harmonics due to symmetry breaking, in agreement with previous findings \cite{Gorlach2020}. At $r_0 = 12020$ nm the Bessel beam has a smaller amplitude in comparison to the hot spot and almost zero gradients (see Fig. \ref{fig:cross_section} b). Less field amplitude naturally results in an earlier cut-off which is found at the 11th harmonic, as well as weaker peaks. While the field gradient effects are minimal, the magnetic field effect causes quadrupolar yield at even harmonics also in this case. Quadrupolar emission increases with higher harmonics, due to its $\omega^6$ dependence (Eq. \ref{eq:quad_contribution}), reaching considerable yields at the hot spot, owing to the larger cut-off. We can deduce the quadrupolar emission (which is fully addressed only within beyond dipole approaches) will become even more prominent at higher intensities which will entail higher harmonics. This yields a fresh perspective confirming the necessity of using beyond electric dipole approximation as the intensity increases \cite{Reiss2008}.

\begin{figure*}
    \includegraphics[scale=0.5]{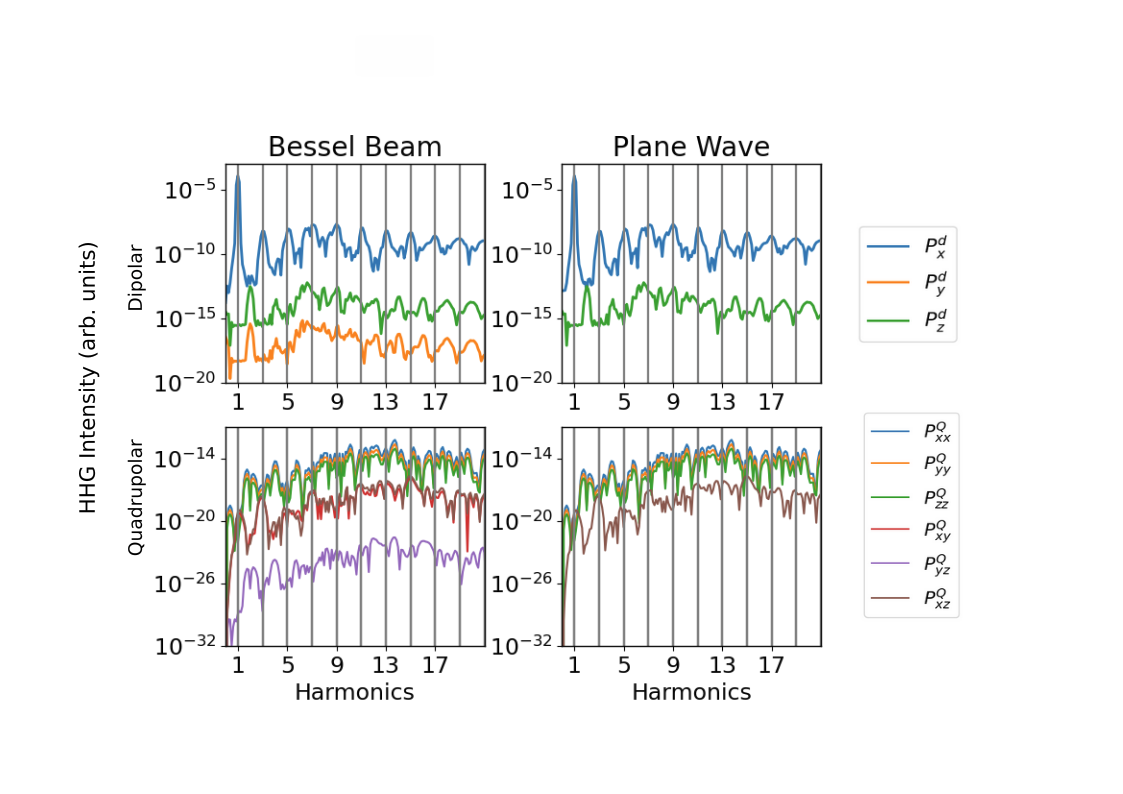}
    \caption{Component resolved spectra for a plane wave beam and Bessel beam for the case when the atom is located at 2545 nm (hot spot). At this point, the field amplitude, frequency, phase, and envelope of the Bessel beam and plane wave match. The upper panel depicts the dipolar emission from all three directions, while the lower panel is reserved for quadrupolar emission for each direction of the quadrupolar tensor. Here, the presented results are calculated beyond dipole approximation.}
    \label{fig:directional_spec}
\end{figure*}
 We now compare the spectra resulting from a Bessel beam and a plane wave to identify the OAM effects. In Fig. \ref{fig:directional_spec}, we show the spectra induced by the Bessel beam on the left and the plane wave spectra on the right.  We display different spatial components and contributions from different multipolar order (dipolar and quadrupolar emission). When the coupling Hamiltonian is in dipole approximation (velocity gauge), the spectra of the Bessel beam and plane wave are indistinguishable (See Supporting Information). This limitation of the electric dipole approximation stems from the inability to capture the spatial dependence of the beam, which is crucial for OAM beams.

 To go beyond this limitation, in Fig. \ref{fig:directional_spec} we show the results in full minimal coupling. The dipolar and quadrupolar spectra reveal that the Bessel beam can excite all dipole and quadrupole components, while the plane wave only excites the dipole in the $x$ and $z$ directions. The dipolar contribution in the $z$ direction, i.e. the direction of propagation for both beams, is a clear result of beyond-dipole coupling \cite{Forre2006,Madsen2022}. While the quadrupolar nature of the Bessel beam provides access to every quadrupolar component, the plane wave is missing the cross components $\tilde{Q}_{yz}(\omega)$ and $\tilde{Q}_{xy}(\omega)$, due to the absence of field gradients in the $y$ direction. For a full view of two set of spectra, please refer to the Supporting Information, where the components with zero yield are not omitted. The fact that both spectra have contributions from components other than the beam polarization ($x$) direction shows the need for a beyond dipole approach to resolve the spectrum thoroughly. 

\begin{figure*}
    \includegraphics[scale=0.25]{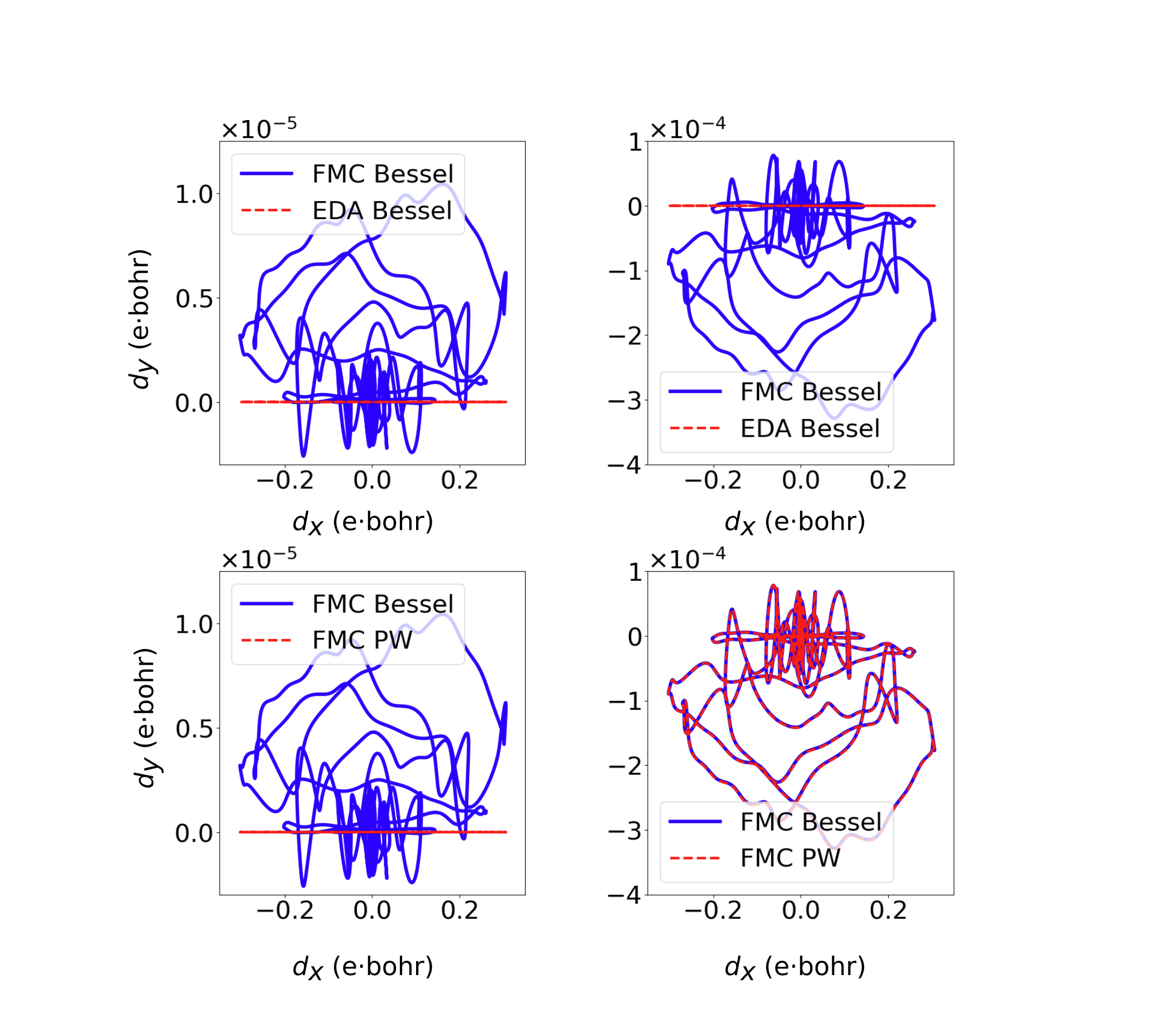}
    \caption{Comparison of dipole moment $d_i(t)$ (See \eqref{eq:dip_moment}) dynamics with Bessel Beam (OAM) and plane wave using electric dipole approximation and full minimal coupling approaches. In the upper part, we provide the dynamics from the Bessel beam in full minimal coupling (FMC Bessel, blue) and velocity-gauge dipole (EDA Bessel, red) approaches. It is visible that the dipole coupling is unable to resolve off-axis deviations. The lower panel compares plane wave-induced off-axis motion and OAM beam-induced motion. The lower right panel shows that the plane wave also has a beyond dipole effect on the electron's trajectory.}
    \label{fig:off_axis_motion}
\end{figure*}

 To unravel the off-axis contributions excited by the two beams, we show the dipole moment $d_i(t)$ \eqref{eq:dip_moment} dynamics in Fig. \ref{fig:off_axis_motion}. To observe the animated trajectory depicted in this figure as a function of time, please refer to animations published in Ref. \cite{animation}. In the upper panels, we depict the trajectory induced by the Bessel beam. Blue lines indicate results in full minimal coupling while red lines depict those in electric dipole approximation. While the latter captures a one-directional dipole oscillation along the polarization direction, the full minimal coupling shows the activation of both $y$ and $z$ dipole components, with corrections in the order of 0.01\% and 0.1\%, respectively.

The off-axis motion can be explained considering the electron trajectory: as the electron travels further away from the ion, it accelerates and reaches a higher velocity. This results in a larger Lorentz force acting on the electron, which in turn causes deviations from a linear path constrained to the polarization axis of the incoming beam \cite{Maurer2021}. Hence, considering only the electric field amplitude (dipolar coupling) when calculating the electron's trajectory is insufficient \cite{Sarachik1970,Reiss2014}. Deviations due to the magnetic field are already visible for both beams in Fig. \ref{fig:off_axis_motion} (lower panels). As expected, they are identical in the $xz$ plane, considering the electric and magnetic field amplitudes, as well as the field gradients in the direction of propagation, of both Bessel beam and plane wave are adjusted to be equal.

As the effects in the propagation direction are equal, we can now isolate the effect of cross-sectional gradients more easily, which only occur in Bessel beams. We observe that in addition to the magnetic-field-induced figure-eight motion in the $xz$ plane, the OAM-carrying Bessel beam induces changes in the electron trajectory in all directions due to the $xy$-plane field gradients, changing the angular dependence of the emission (\textit{vide infra}) and exciting all of the quadrupole moment components. The fingerprint of such an effect is visible in Fig \ref{fig:off_axis_motion} (lower left panel), evidenced by oscillations in the $y$ axis which are absent in the case of the plane wave. The $y$-dipole trajectory gives then rise to the $P^d_y$, $P^Q_{xy}$ and $P^Q_{yz}$ HHG spectra in Fig. \ref{fig:directional_spec}.

\begin{figure*}
    \includegraphics[scale=0.45]{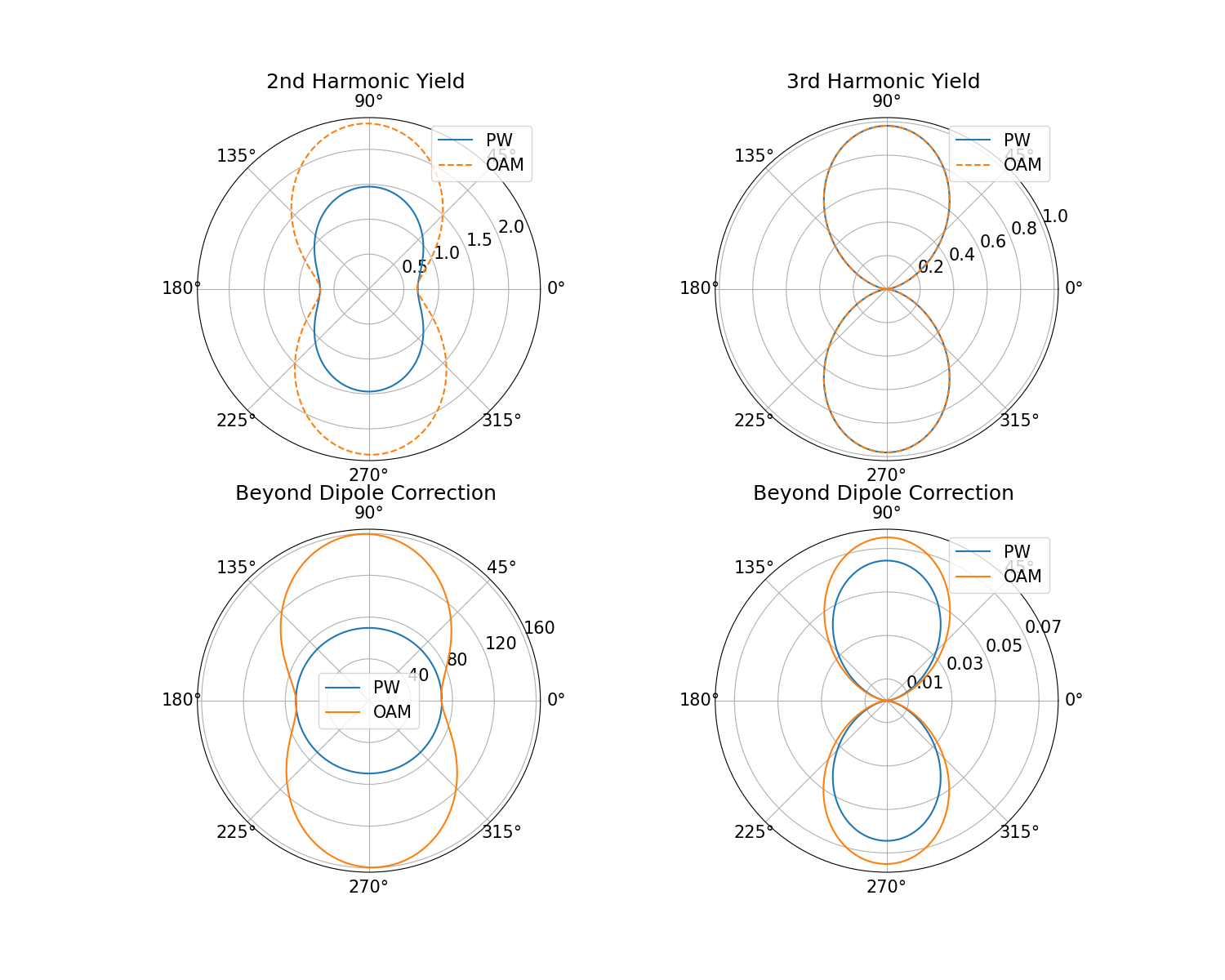}
    \caption{Angular dependence in the $xy$ plane of the emitted power $\textbf{P}^d(\omega, \theta)$ for second and third harmonics. The panels on the left depict the results from the second harmonic while the ones on the right show the results from the third harmonic. The upper panel from the left describes the total emission induced by a plane wave (PW) and Bessel beam  while the lower left one describes the full minimal coupling correction compared to the dipole approximation. Similarly, the right side shows total emission induced by a plane wave (PW) and Bessel beam for the third harmonic on the upper part, and the beyond dipole correction on the lower panel. Results here are normalized by total emission $P_{\mathrm{pw}}^d(\omega)= \int r^2 d\Omega P^d_{\mathrm{pw}}(\omega, \theta)$, where $d\Omega = \sin\theta d\theta d\varphi$ induced by the plane wave.}
    \label{fig:emission_char}
\end{figure*}

 Now we show the beyond dipole correction on the angular distribution of the resulting emission. As the main observable of the HHG process is the yield of harmonics, we calculate the emitted power, see Fig. \ref{fig:emission_char}. The angle-dependent power dissipation $P^d(\omega, \theta)$ is given by
\begin{equation}
    P^d(\omega, \theta)  = \frac{\mu_0 |\tilde{\boldsymbol{d}} (\omega)|^2 \omega^4}{32 \pi^2 c r^2}  \sin^2\theta.
    \label{eq:power_angle}
\end{equation}
Here, $\theta$ is the angle between dipole moment $\tilde{\boldsymbol{d}} (\omega)$ and the position vector $\bf{r}$. For our calculations we assume $|\bf{r}| = 1 $ $\mu$m. For visualization, the expression is then normalized by the total power $P_{\mathrm{pw}}^d(\omega) = \int r^2 d\Omega P_{\mathrm{pw}}^d(\omega, \theta)$, where $d\Omega = \sin\theta d\theta d\varphi$ is the solid angle.
 
 In Fig.~\ref{fig:emission_char}, we present dipole emission of the second and third harmonics in the top left and top right panels, respectively, resolved in azimuthal angle residing in the $xy$-plane (orthogonal to the propagation direction of the beams). Additionally, we present the angular-resolved beyond-dipole corrections in the $xy$ plane. The emission resulting from both the plane wave and the Bessel beam is seen in the upper left panel, calculated in the full minimal coupling. The difference in the angular dependence of the normalized $P^d(\omega, \theta)$ for the Bessel beam and the plane wave is larger in the case of even harmonics, as evidenced by the second harmonic. As can be seen in the upper left panel, the amplitude of emission due to the OAM beam is twice that of the plane wave along the $y$ axis (90$^\circ$ and 270$^\circ$). 
By inspecting the difference of emission calculated within dipole approximation from that in the full minimal coupling approach (beyond dipole correction, lower left panel), it is clear that the angular difference is due to field gradient effects (off-axis motion), reaching up to 160 \% difference in amplitude and a slight tilt due to the $P_y$ component. A correction for the plane wave case is still present due to the magnetic field effects along the $z$ axis, hence the correction has no angular dependence in the $xy$ plane (see Fig. \ref{fig:off_axis_motion}).

The difference in the distribution becomes much smaller for the odd harmonics, as seen in the case of the third harmonic angular spectrum (upper right panel). Here, the dipolar contribution dominates heavily and renders the beyond dipole correction small (on the order of 0.07 \%, as seen in the beyond dipole correction, lower right panel). Hence, the plane wave correction in the second harmonic is around $10^3$ times larger than that of the third harmonic.

Observing the large corrections in even harmonics is exciting for experimental realizations. Such corrections entailed by OAM beams have been reported for large odd harmonics before \cite{albaXtremeUV}. Added to the possibility of measuring even harmonics without dipolar background \cite{Jensen2022}, we believe the considerable correction in the emission visible in Fig. \ref{fig:emission_char} due to the deviations in electron trajectory is also a fitting candidate for an experimental observable, which provides a playground where the effect of OAM can be used to induce emission in all directions.

\begin{figure*}
    \includegraphics[scale=0.45]{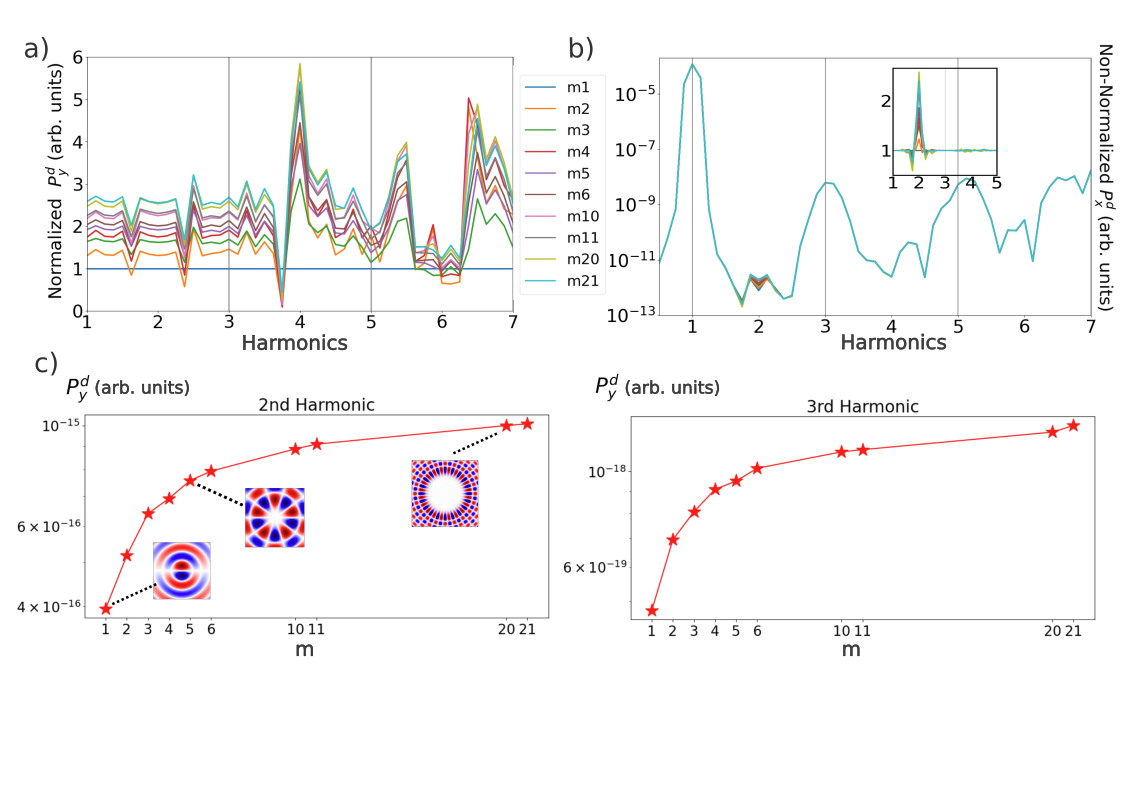}
    \caption{Spectra as a function of OAM number $m$. Here, panel (a) shows the $y$ component of dipolar yield $P^d_y$ for $m$ ranging from 1 to 21, normalized by the spectrum for the $m=1$ case. It is visible that as the $m$ varies, the $y$ component of the emission is adjusted accordingly. Meanwhile, panel (b) depicts the $x$ component $P^d_x$ where the yield is not normalized. Here, the deviation caused by different $m$ is mostly located at the second harmonic. Panel (c) shows the $y$ component yield for the range of $m$ for second and third harmonics, respectively, showing approximately a saturation regime after $m=10$. Here, the inset in the second harmonic graph displays the cross sections of beams with $m=1$, $m=5$, and $m=20$. In both panels of part (c), odd $m$'s are marked with red, and even ones are marked with blue color. Here, the presented results are calculated beyond dipole approximation.}
    \label{fig:m_s}
\end{figure*}

 Finally, to investigate the effect of orbital angular momentum number, we construct linear Bessel beams with higher $m$. We place the atom at the first intensity ring for each case of $m$, with the amplitude and phase of the beams matched, so that the dipolar effects are identical for all the beams.
 
By doing this, the beyond-dipole effects are isolated and the trends can be studied without a dipolar background. We investigate values of $m$ ranging from 1 to 21. In Fig. \ref{fig:m_s}, panel (a), we show dipolar $y$ spectra $P^d_y$ normalized by the $m=1$ case, for different OAM numbers, which evidences that the emission can be enhanced by tuning the incident OAM. Particularly up until the 4th harmonic, the increase of yield is monotonic with the OAM number. In panel (b) of Fig. \ref{fig:m_s}, the (non-normalized) emission in the $x$ direction $P^d_x$ is shown, while the inset depicted the normalized values with respect to the $m=1$ case. Even in this direction, where the dipolar mechanism driven by the amplitude of the laser is dominant, the apparent deviation in the second harmonic shows that beyond-dipole effects due to the OAM are visible in the full spectrum. However, the main deviation in the $x$ direction is found in the second harmonic, in contrast to the $y$ direction where the effect of $m$ is present in a range of harmonics.

Differences between the $m$ dependence in different directions can be attributed to larger trajectories along the $x$ axis, resulting in the electron experiencing field gradients further away from the hot spot in this direction.

Panel (c) of Fig. \ref{fig:m_s} shows the $y$ component of the yield as a function of OAM number $m$ for the second (left) and third (right) harmonics. While the yield increases as a function of $m$ for both cases, a saturation regime is reached for larger values of $m$. This can be understood by the field gradient following the same trend for increasing values of $m$, as shown in the Supporting Information for the $y$-derivative of the electric field at the hot-spot position of each beam. This finding shows that the nonlinear off-axis response of the system depends on the OAM number of the beam following a similar trend to the field gradients, which provides a path to include beyond-dipole effects in simulations of HHG from the gas-phase target.

\section{Conclusion}

\noindent To conclude, we identify beyond-dipole effects in HHG in the presence of OAM beams. We predict that such corrections are particularly relevant for even harmonics, which appear due to symmetry breaking. We show that off-axis trajectories are well captured beyond the dipole and modify considerably the angular distribution of the emitted power for even harmonics.
A consequence of using spatially-structured light is the deviation of the electron trajectory in the axis orthogonal to the propagation and polarization directions, caused by the field gradients in the plane perpendicular to the propagation axis. While this off-axis motion is small compared to the motion in the direction of propagation, the corrections could be substantial, especially for even harmonics. Finally, we show that the harmonic yield in both off-axis and on-axis directions can be enhanced by tuning the OAM number of the Bessel beam, where the corrections follow the trend of the electric field gradients. Future calculations of propagation in a medium, accounting both for microscopic and macroscopic effects, will provide further insight into the experimental setups to detect and harness the beyond-dipole effects for nonlinear spectroscopy with structured light.

\section*{Data Availability Statement}
Raw data were generated at the Max Planck Computing and Data Facility using open source code Octopus-code - version $15.0$ . Derived data supporting the findings of this study are available from the corresponding author upon reasonable request.

\section*{Conflict of Interest}
The authors declare no conflicts of interest relevant to the content of this article.

\begin{acknowledgments}

The authors also would like to acknowledge the computational support provided by Max Planck Computing and Data Facility. We are grateful for the help of the Octopus Developers team, and we thank Dr. Nicolas Tancogne-Dejean, Dr. Simon Vendelbo Bylling Jensen, and Dr. Alba de las Heras for valuable discussions. This work was supported by the European Research Council (ERC-2015-AdG694097), the Cluster of Excellence ‘Advanced Imaging of Matter’ (AIM), Grupos Consolidados (IT1453-22) and Deutsche Forschungsgemeinschaft (DFG) - SFB-925 - project 170620586. The Flatiron Institute is a division of the Simons Foundation. We acknowledge support from the Max Planck-New York City Center for Non-Equilibrium Quantum Phenomena. E.I.A. acknowledges support from International Max Planck Research School. F.P.B. acknowledges financial support from the European Union’s Horizon 2020 research and innovation program under the Marie Sklodowska-Curie Grant Agreement no. 895747 (NanoLightQD).

\end{acknowledgments}
\bibliography{apssamp}

\appendix 
\title{Supporting Information: High Harmonic Generation with Orbital Angular Momentum Beams: Beyond-dipole Corrections}

\author{Esra Ilke Albar}
\email{esra-ilke.albar@mpsd.mpg.de}
\affiliation{ 
Max Planck Institute for the Structure and Dynamics of Matter, Center for Free Electron Laser Science, Luruper Chaussee 149, 22761 Hamburg, Germany
}

\author{Valeriia P. Kosheleva}%
 
\affiliation{ 
Max Planck Institute for the Structure and Dynamics of Matter, Center for Free Electron Laser Science, Luruper Chaussee 149, 22761 Hamburg, Germany
}

\author{Heiko Appel}%
\affiliation{ 
Max Planck Institute for the Structure and Dynamics of Matter, Center for Free Electron Laser Science, Luruper Chaussee 149, 22761 Hamburg, Germany
}%

\author{Angel Rubio}%
\affiliation{ 
Max Planck Institute for the Structure and Dynamics of Matter, Center for Free Electron Laser Science, Luruper Chaussee 149, 22761 Hamburg, Germany
}%
\affiliation{%
Center for Computational Quantum Physics (CCQ), The Flatiron Institute, 162 Fifth Avenue, New York, New York 10010, USA
}

\author{Franco P. Bonafé}
\email{franco.bonafe@mpsd.mpg.de}
\affiliation{%
Center for Computational Quantum Physics (CCQ), The Flatiron Institute, 162 Fifth Avenue, New York, New York 10010, USA
}

\date{\today}

\maketitle

\section{Supporting Information and Additional Notes}
\subsection{Construction of linearly polarized Bessel beam} \label{construct_linear_bessel}
Similar to the case of conventional plane-wave light beams linearly polarized Bessel beam can be constructed as follows:  
\begin{equation}
\boldsymbol{A}_x^{(\mathrm{tw})}(\mathbf{r},t)
\approx\frac{1}{\sqrt{2}}\left[A_{m_{\gamma_1}, \lambda=+1, \theta_k\rightarrow 0,x}^{(\mathrm{tw})}(\mathbf{r},t)+A_{m_{\gamma_2}, \lambda=-1, \theta_k\rightarrow 0,x}^{(\mathrm{tw})}(\mathbf{r},t)\right]\boldsymbol{e}_x,
\label{superposition}
\end{equation}
where $m_{\gamma_1}-m_{\gamma_2} = 2$, $m_{\gamma}$ is total angular momentum, $\lambda = \pm 1$ is a helicity, $\theta_k = \arctan\left( \frac{\varkappa}{k_z} \right)$ is so-called opening angle, and $A^{(\rm tw)}_{m_{\gamma}, \lambda, \theta_k,x}(\mathbf{r},t)$ is an $x$ component of the vector potential,
\begin{equation}
\begin{aligned}
A^{(\rm tw)}_{m_{\gamma}, \lambda, \theta_k,x}(\mathbf{r},t) & =\frac{A_0}{\sqrt{2}}\left[\cos \left(k_z z+\phi(m_{\gamma}+1)-\omega t\right) d_{-1 \lambda}^1\left(\theta_k\right) J_{m_{\gamma}+1}\left(\frac{\varkappa r_{\perp}}{c}\right)\right. \\
& \left.+\cos \left(k_z z+\phi(m_{\gamma}-1)-\omega t\right) d_{1 \lambda}^1\left(\theta_k\right) J_{m_{\gamma}-1}\left(\frac{\varkappa r_{\perp}}{c}\right)\right]. 
\end{aligned}
\end{equation}
Here $d_{\mu \lambda}^1\left(\theta_k\right)$ is a small Wigner matrix given by
\begin{equation}
d_{\mu \lambda}^1\left(\theta_k\right) = \begin{cases}\frac{\sin \theta_k}{\sqrt{2}} \lambda, & \text { if } \mu=0 \\ \frac{1+\lambda\cos \theta_k }{2}, & \text { if } \mu=1 \\ \frac{1-\lambda\cos \theta_k }{2}, & \text { if } \mu=-1\end{cases}
\end{equation}
One can show that Eq. \eqref{superposition} indeed resembles the equation in the main text using the relation that $\sin \left(\theta_k\right) \approx \theta_k$ and $\cos \left(\theta_k\right) \approx 1$ and $m_{\gamma} = m + \lambda$, where $m$ is OAM number  for $\theta_k\rightarrow0$.
We use Eq. \eqref{superposition} with $\theta_k = 5^{\circ}$ in the current investigation.

\subsection{Directional Spectra: Further Analysis}
\begin{figure*}
    \includegraphics[scale=0.5]{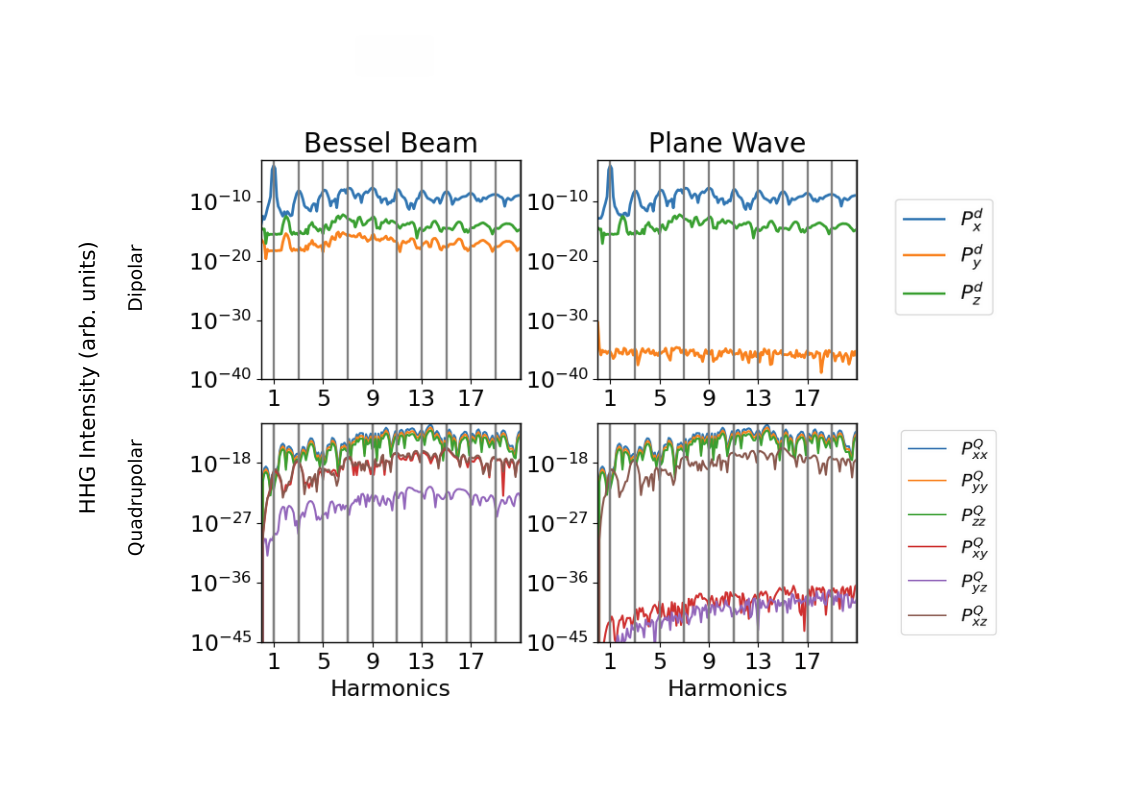}
    \caption{Directionally resolved spectrum for both cases, calculated beyond dipole approximation. Zero contributions are included to highlight the excitation provided by the Bessel beam. }
    \label{fig:full_fmc_spec}
\end{figure*}

\begin{figure*}
    \includegraphics[scale=0.5]{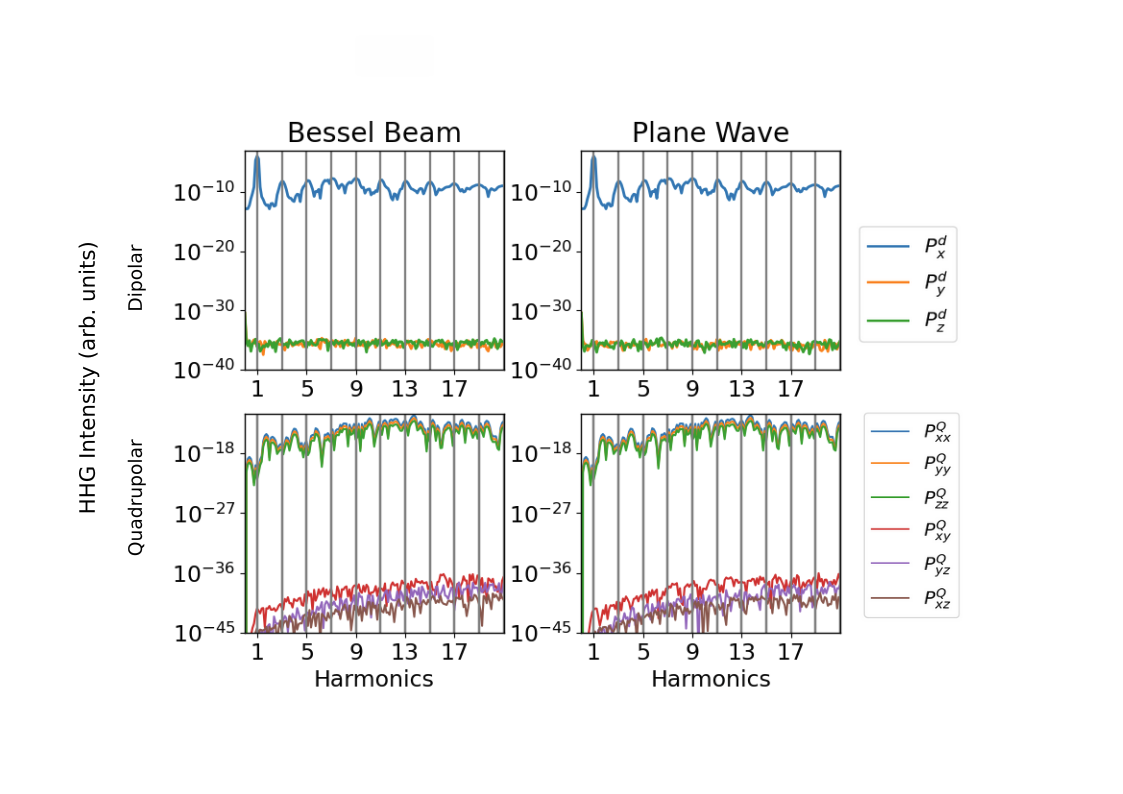}
    \caption{Directionally resolved spectrum for both cases, calculated in velocity dipole approximation. The emission spectrum is shown in terms of harmonics. The spectra of the Bessel beam and plane wave are not distinguishable within the dipole approximation. }
    \label{fig:vd_directional_spec}
\end{figure*}

Here, detailed full minimal coupling spectra (Fig. \ref{fig:full_fmc_spec}), along with velocity gauge dipole spectra (Fig. \ref{fig:vd_directional_spec}) are presented.

\subsection{Field gradient dependence with OAM number}
\begin{figure*}
    \includegraphics[scale=0.5]{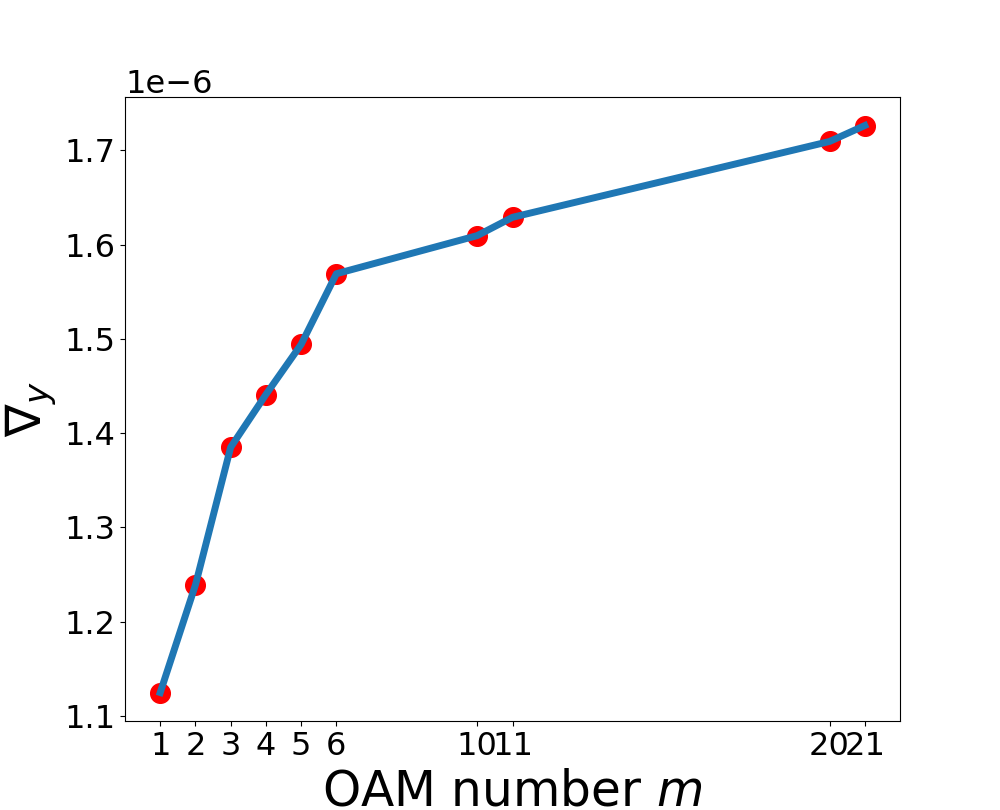}
    \caption{ Maximum $y$ gradient ($\nabla_y$) for each OAM number $m$. The rapid initial increase of the amplitude of the gradient is saturated as one moves to higher OAM numbers.  }
    \label{fig:grad_y}
\end{figure*}

Fig. \ref{fig:grad_y} shows the magnitude of $y$ direction gradient at the hotspots of each $m$ number.

\end{document}